**Title:** Population genetics of rare variants and complex diseases

**Authors:** M. Cyrus Maher[a+], Lawrence H. Uricchio[b+], Dara G. Torgerson[c], and Ryan D. Hernandez[d*].

[a]Department of Epidemiology and Biostatistics, University of California, San Francisco.

[b]UC Berkeley & UCSF Joint Graduate Group in Bioengineering, University of California, San Francisco.

[c]Department of Medicine, University of California San Francisco.

[b]Department of Bioengineering and Therapeutic Sciences, University of California, San Francisco.

[+]These authors contributed equally.

[*]Corresponding author:

Ryan D. Hernandez, PhD.

Department of Bioengineering and Therapeutic Sciences

University of California San Francisco

1700 4[th] St., San Francisco, CA 94158.

Tel. 415-514-9813, E-Mail: ryan.hernandez@ucsf.edu

**Abstract**:

**Objectives:** Identifying drivers of complex traits from the noisy signals of genetic variation obtained from high throughput genome sequencing technologies is a central challenge faced by human geneticists today. We hypothesize that the variants involved in complex diseases are likely to exhibit non-neutral evolutionary signatures. Uncovering the evolutionary history of all variants is therefore of intrinsic interest for complex disease research. However, doing so necessitates the simultaneous elucidation of the targets of natural selection and population-specific demographic history. **Methods:** Here we characterize the action of natural selection operating across complex disease categories, and use population genetic simulations to evaluate the expected patterns of genetic variation in large samples. We focus on populations that have experienced historical bottlenecks followed by explosive growth (consistent with most human populations), and describe the differences between evolutionarily deleterious mutations and those that are neutral. **Results:** Genes associated with several complex disease categories exhibit stronger signatures of purifying selection than non-disease genes. In addition, loci identified through genome-wide association studies of complex traits also exhibit signatures consistent with being in regions recurrently targeted by purifying selection. Through simulations, we show that population bottlenecks and rapid growth enables deleterious rare variants to persist at low frequencies just as long as neutral variants, but low frequency and common variants tend to be much younger than neutral variants. This has resulted in a large proportion of modern-day rare alleles that have a deleterious effect on function, and that potentially contribute to disease susceptibility.

**Conclusions:** The key question for sequencing-based association studies of complex traits is how to distinguish between deleterious and benign genetic variation. We used population genetic simulations to uncover patterns of genetic variation that distinguish these two categories, especially derived allele age, thereby providing inroads into novel methods for characterizing rare genetic variation driving complex diseases.

**Introduction**:

Over the last 12,000 years, humans have experienced explosive population growth, increasing in number by three orders of magnitude. Such extreme growth has caused the dynamics of mutation accumulation and the distribution of allele frequencies to diverge significantly from the patterns expected at equilibrium. Moreover, mutations arise every generation at a rate of 1.1-3 $\times 10^{-8}$ per base in the human genome (*1-3*). Since there are ~$3\times10^9$ bases in the human genome, we expect a given person to have, on average, between 30-100 *de novo* mutations [consistent with observations in large family-based sequencing projects (*1*)]. With ~130M people born each year, it is estimated that every base in the human genome has the potential to mutate 1-4 times every year. Therefore, we expect every non-lethal mutation that is possible in the human genome to be present in numerous individuals throughout the world, and potentially on distinct haplotype backgrounds.

Following the introduction of a germline mutation, the evolutionary forces of genetic drift and natural selection push the derived allele either towards loss from the population, or towards increasing frequency and possible fixation. If an allele is subject to natural selection, then large genomic areas can also be impacted due to genetic linkage. In the case of strong positive selection, a region of the genome can be swept free of genetic variation. This scenario, referred to as a "selective sweep", appears to have been rare in recent human evolution (*4*). In the case of purifying selection, when deleterious alleles are recurrently purged from a locus, neutral variation in linked genomic neighborhoods can also be reduced. This much more common scenario is termed "background selection" (*5, 6*). Signatures of both positive and negative selection have been identified throughout the human genome, but the relevance of identifying such signatures for genetic studies of human diseases has not been well studied.

Characterizing how neutral and selective evolutionary processes have impacted modern day patterns of human genetic variation is an important step toward understanding the genetic drivers of disease. However, this requires extremely large-scale sequencing data. To date, thousands of human genomes (and thousands more exomes) have been sequenced (*7-9*). These large-scale efforts have resulted in a flurry of discoveries about the patterns of genetic variation within and between human populations. However, a mechanistic understanding about how the observed patterns of genetic variation impact human traits is still lacking.

The sequencing of a small number of genes in more than ten thousand humans from a single population identified far more variants than previously expected (*10, 11*). Population genetic theory suggests that exploding human population growth will result in an accumulation of extremely rare variants. We hypothesize that some portion of this rare, potentially deleterious variation underlies much of the unexplained heritability of many complex traits (*12*).

Recent sequencing of thousands of exomes from across the world revealed that the vast majority of genetic variation within genes is rare, arose recently, and is highly population-specific (*7-9, 13*). Whole-genome sequencing of more than 1000 individuals from global populations has elucidated the broad-scale patterns of genetic variation in much greater detail (*7*). Such large-scale data suggest that human evolutionary history is more complex than originally thought (*4*). These observations are consistent with population genetic theory, and suggest that population genetics may be useful for elucidating the major genetic drivers of many complex diseases (*14*).

The next phase of human genetic research will sequence ever-larger cohorts in order to pinpoint the genetic contributors to complex disease. Such directions will undoubtedly identify an ever-increasing number of candidate variants that require follow-up. Population genetic theory may be of great utility for prioritizing candidate functional variants, thus allowing for a more efficient allocation of experimental resources.

In this paper, we will demonstrate how population genetics can inform expectations about the nature of undiscovered disease-causing variants, and accelerate experimental discovery. In addition, we will show that the unique genetic signatures that we expect from deleterious alleles compared to other variants may increase the power of statistical tests of association. Although population genetic models have been developed for several decades to characterize the expected signatures of genetic variation driven by natural selection, such models have rarely been employed in genetic studies of complex disease [though see (*15, 16*)]. We will demonstrate that this is a promising avenue for future development.

We will accomplish this through simulations, and through the analysis of genomic data. We begin our investigation by looking at negative selection across the exome. If complex disease-causing mutations were under the influence of natural selection, we would expect to observe evolutionary signatures in the genes harboring those variants, even in a sample of healthy individuals. Previous studies have identified evidence for stronger negative selection on Mendelian disease genes (*17, 18*), and noted a similar trend in candidate *cis*-regulatory regions of genes associated with complex diseases as a whole (*18*). Here, we extend these analyses to test for an enrichment of negative selection across genes associated with specific categories of complex diseases by analyzing resequencing data from the exomes of 20 European Americans and 19 African Americans (*17, 18*). Similarly, we also test for intensified background selection at loci previously associated with complex disease through GWAS.

In addition, we perform extensive simulations of deleterious variation under a complex demographic model that incorporates recent explosive population growth to describe characteristic features of deleterious variation. Overall, our results demonstrate that population genetic approaches can contribute greatly to the development of novel methods to identify causal genetic variation that contributes to disease. We expect that methods that integrate experimental data with population genetic theory will provide a great advantage in the hunt for the "missing heritability" of complex disease.

**Methods**

*Disease association datasets*

We used two disease association databases in this study, including the Genetic Association Database [GAD (*19*)] for our analysis of natural selection near genes associated with disease, and the catalog of GWAS targets published by the NHGRI (*20*) for our analysis of background selection at GWAS loci. The latter set has also been annotated into specific disease/trait categories (*21*). These two resources enable complementary analyses, since the GAD provides a gene-centric view of complex disease including candidate genes, while the NHGRI GWAS catalog specifically focuses on genome-wide studies.

*Measuring natural selection on disease genes*

Most classic statistical tests for natural selection are built on the idea of comparing two categories of mutations: those subject to natural selection (e.g., non-synonymous mutations) and those that are not (e.g., synonymous mutations). From a population genetic perspective, the action of natural selection is mainly to increase or decrease the probability of a mutant allele becoming fixed in a population. The McDonald-Kreitman (MK) Test (*22*) is a powerful statistical test for natural selection that is based on this principle. This test compares the number of mutations that have become fixed between species to the number that are polymorphic within a population across synonymous and non-synonymous sites in a two-by-two contingency table.

Torgerson *et al.* (*18*) developed a Bayesian extension of the MK test for estimating the probability that negative selection was operating on each gene in the genome for both coding regions (CDS; comparing non-synonymous and synonymous sites) and conserved non-coding regions (CNC; comparing CNC sites to synonymous sites). The approach was applied to exome sequence data from 20 European American individuals and 19 African American individuals, and we used their results in our study.

We applied a Mann-Whitney *U* test to compare the distribution of the probability of negative selection in genes associated with 17 classes of complex disease from the Genetic Association Database (*19*) to the distribution of genes not previously associated with complex disease. We then perform a binomial test to evaluate whether there is an enrichment of significant tests at a nominal level of $p<0.05$ for each population and sequence annotation (CDS versus CNCs), in which we used 0.05 as our theoretical probability of success and the total number of complex disease/trait categories that fell into each bin of Table 1.

*Background Selection Scores*

Diversity at a neutral locus can be reduced by selection acting at genetically linked loci in a process referred to as background selection (*4-6, 23-*

*26*). The effect of background selection was recently inferred for each base of the human genome (*25*). We transformed the scores, *B*, to range from 0 (no effect of background selection) to 1 (diversity completely reduced due to background selection). Fractional values between zero and one represent the relative reduction in diversity due to background selection, compared to what is expected at a neutral locus (e.g., *B*=0.5 indicates that background selection has reduced diversity by 50%). For each GWAS SNP, we extracted the *B* score from (*25*), the allele frequency in the CEU population of HapMap, and the genetic distance to the nearest gene in hg19.

*Choosing control SNPs for GWAS analysis*

One thousand sets of control SNPs were chosen via rejection sampling. Each set was matched in distribution to GWAS SNPs based on allele frequency, and in the case of intergenic SNPs, genetic distance to the nearest gene. Allele frequencies were taken as the CEU HapMap frequency, and for each GWAS SNP, a centered window of 0.25cM was excluded from sampling. Average background selection scores were then calculated for each base in the control set. We then generated a quantile-quantile plot of the effect of background selection at GWAS loci versus the expected effect of background selection given our control sets (taken as the mean of the order statistics across the 1000 sets). A confidence interval was estimated as the entire range of the order statistics across all 1000 sets.

*Diversity versus phyloP*

PhyloP scores are a measure of evolutionary conservation generated from a multispecies alignment. For our study, we used an alignment of 46 placental mammals (*27*). Positive phyloP scores indicate conservation across species while negative values indicate rapid evolution across species. PhyloP scores were obtained from the UCSC genome browser (http://hgdownload.cse.ucsc.edu/goldenPath/hg19/phyloP46way/). We calculated per site diversity for CEU, YRI and CHB in the 1000 Genomes Project Phase I data set (*7*). Diversity was calculated as 2*p* (1 - *p*) *N*/(*N*-1), where *p* is the allele frequency and *N* is the number of chromosomes sequenced in the population. We have previously shown that, despite the low coverage nature of the genome-wide sequencing data set, this estimator accurately recapitulates diversity when a sufficient number of sites are included in the calculation (*4*). Sites across the genome with similar phyloP scores were binned in intervals of 0.08, and mean diversity was calculated for each bin. The data were fit to a linear model relating $\log(\pi)$ ~ phyloP. Only sites with phyloP scores between -1.2 and 1.2 were used to fit the model, and the regression line was extrapolated across the full range of phyloP scores (-4, 3).

*Distribution of selection coefficients*

Natural selection acts to increase the probability that advantageous mutations increase in frequency, while constraining the frequency of deleterious mutations. In this study, we were primarily concerned with the effects of deleterious mutations. For the simulations discussed below, we used the distribution of selection coefficients inferred previously from an exome-scale sequence data set (*28*). The parameters for the distribution of selection coefficients were inferred for both European Americans as well as African Americans, and the results were largely concordant. We therefore incorporated the previously published distribution directly, which is highly leptokurtic with: $-s \sim \Gamma(0.184, 0.00040244)$. Note that we only consider deleterious selection coefficients ($s<0$), and will sometimes report the absolute value of the selection coefficient for convenience. Also note that the same individuals that were used to estimate the distribution of selection coefficients were also used to measure the probability of natural selection acting on disease versus non-disease genes above.

*Demographic history*

We used the demographic history inferred from exome data sequenced across 2,440 samples from two populations (*9*). This model incorporates the demographic history of both a European and African population. Our simulations were focused on a European demographic model, but are relevant for all populations that have experienced historical bottlenecks and recent growth. We focused on this population, as the vast majority of GWAS findings are from European populations (*29*), and for the most part it remains to be seen whether the same loci are associated in non-European populations.

Extracting the European demographic history from the two-population model may lead to systematic differences in the allele frequency spectrum due to the absence of historical migration. However, such events are unlikely to substantially contribute to rare genetic variation, as recent migration rates are low and the age of rare variants is too short to have arisen prior to the Out Of Africa event. The demographic history included an ancient African expansion (~177kya), an out-of-Africa bottleneck (~62kya), a founding of Europe bottleneck (~28kya), an initial phase of exponential growth within Europe, and a recent explosive growth phase (starting ~5kya).

*Simulations*

All simulations were performed using SFS_CODE (*30*). This is a flexible forward time population genetic simulator that is capable of handling arbitrarily complex demographic histories and distributions of selection coefficients. We performed $10^6$ simulations of a gene of length 1500bp, including the model of human demographic history and distribution of selection coefficients operating on non-synonymous mutations (discussed above). At the beginning of each simulation, a random nucleotide sequence was generated, and a population with 2,000 diploid individuals was created. Random mating occurred for a burn-in period of 20,000 generations to reach mutation-selection balance, at which point the demographic model ensued. We then sampled 5,000 diploid individuals (drawn from the terminal population of size 140,086 individuals). Each generation consisted of random mating between males and females, with a Poisson number of new mutations entering the population each generation (with mean θ/2=0.0005 per base). Generations were converted to years by rescaling to the ancestral population size of 7310 [inferred by (*13*)] and assuming 30 years per generation. A smaller number of simulations were performed using the ancestral size of 7310 individuals inferred by (*13*), with completely concordant results [as expected since the relevant parameters are scaled by the effective population size; see SFS_CODE user manual for longer discussion on this topic (*30*)]. A representative SFS_CODE command line is as follows: `sfs_code 1 1 -A -n 5000 -N 2000 -L 1 1500 -I -r 0.001 -W 2 0 0 0 0.184 0.00040244 -Td 0 1.98 -Td 0.264196 0.12859 -Tg 0.340566 44.75089 -Td 0.340566 0.554513 -Tg 0.39085 282.7192 -TE 0.404845`.

## Results

*Natural selection in regions associated with complex traits*

By analyzing the exomes of 20 individuals from two populations (European American and African American), we find that genes associated with six categories of complex traits are enriched for signatures of negative selection. These include cancer, cardiovascular, developmental, metabolic, and pharmacogenomic traits (Table S1). The number of disease categories that are significant at the 0.05 level are shown in Table 1, with a binomial p-value suggesting an enrichment of negative selection in coding (CDS) regions for both European and African Americans, and in conserved non-coding (CNC) regions only in European Americans for genes associated with complex traits. This suggests that negative selection inferred from population-based samples may be useful for disentangling the genetic basis of complex traits.

Natural selection acting on specific loci throughout the human genome also impacts patterns of genetic diversity in genetically linked regions through an effect known as background selection (*5, 6, 23-26*). We find that the loci that have been significantly associated with complex traits [as identified in the National Human Genome Research Institutes catalog of genome-wide association studies (*20*)] show elevated levels of background selection compared to 1000 control sets of SNPs (each set having the same number of SNPs, allele frequency distribution, and genetic distance to nearest coding gene as the GWAS hits). In Figure 1, we show a quantile-quantile (QQ) plot of the effect of background selection across GWAS SNPs versus the expected effect of background selection at control loci (with the full range of *B* scores for each quantile across all 1000 control sets shown in pink). Using a Mann-Whitney U test to compare the distribution of B scores for GWAS SNPs versus B scores for control SNPs confirms the visual presentation in Figure 1 ($p=1.4\times10^{-138}$).

To evaluate whether the enriched signature of background selection was due to particular types of diseases, we performed the same analysis for loci associated with each of 15 complex diseases and traits (*21*). Figure S1 shows QQ-plots for each disease category, which demonstrates that the enrichment of background selection for GWAS hits is most extreme for seven categories, though all but two [parasitic bacterial disease (N=27) and viral disease (N=99)] are suggestive.

In summary, genes associated with complex diseases and traits are enriched for signatures of negative selection, and genomic loci associated with complex diseases and traits exhibit increased levels of background selection. Together, this suggests that directly utilizing signatures of natural selection can aid in discovering novel loci associated with complex disease.

*Long-term evolutionary trends and patterns of diversity*

In addition to patterns of diversity within human populations, long-term signatures of natural selection can also be obtained by analyzing the genomes of distantly related species (*27*). This information can be used in tandem with population genetic signatures to better identify targets under selection. In Figure 2, we correlate diversity with phyloP [a measure of evolutionary conservation with positive values indicating higher conservation (*27*)]. While long-term evolutionary constraint likely contributes to this correlation, a regression model fit using only neutral values (phyloP between -1.2 and 1.2) and extrapolated to extreme values yields decent performance in predicting the levels of diversity in the most highly constrained regions of the genome (phyloP > 2; though there is an excess of diversity observed across populations) as well as in the most quickly evolving regions of the genome (phyloP < -2).

We then compared the distribution of phyloP scores at the GWAS SNPs discussed above to sets of control SNPs. In contrast to a previous comparison that did not account for frequency or genetic distance to genes (*31*), we find no statistical difference between the two matched groups using a Mann-Whitney U test (p=0.052). This suggests that GWAS SNPs are likely not causal, but, rather, are genetically linked to the causal allele (concordant with our observation of highly enriched signature of background selection around GWAS SNPs).

*The impact of natural selection and demography on rare variation*

Natural selection acts to increase the probability that adaptive mutations will rise in frequency and that deleterious mutations will decrease in frequency. Demographic effects simply modulate the variance in allele frequency change over generations with no bias toward increasing or decreasing frequency. However, when acting simultaneously, deleterious alleles can sometimes rise in frequency (*32*). We used simulations to better understand the composition of fitness effects acting on variants across the frequency spectrum. In Figure 3, we show simulation results using SFS_CODE (*30*) that demonstrate the distribution of fitness effects for variants in several frequency ranges (see Methods for details). While over 30% of all variants that were private to a single chromosome out of a sample of 10,000 chromosomes (5,000 individuals) were moderately to strongly deleterious, only a small fraction of variants with frequency >5% had moderately negative fitness effects. Importantly, variants with frequency >10% had only weak fitness effects.

In addition to having a wide range of fitness effects, variants that have the same frequency can also have a very broad distribution of ages (i.e., the time since the derived mutant allele arose in the population). The shape of the distribution of ages has a critical relationship with the population's demographic history. The demographic history we use is shown in the background of Figure 4 (left axis), which incorporates an ancient expansion, an out-of-Africa bottleneck, a founding-of-Europe bottleneck, and two phases of exponential growth (an early, slower phase, and a more recent explosive growth phase; see Methods). Figure 4 also shows violin plots for the ages of non-synonymous mutations in various frequency ranges (singletons, doubletons, <0.01%, etc.; right axis).

Consistent with previous observations (*13*), we find for these simulations that essentially all non-synonymous variants that are <1% frequency will have a negligible probability of being old enough to be shared across continents without requiring high rates of migration (i.e., they have arisen since the founding-of-Europe bottleneck). This has profound implications for studying the genetics of complex traits across populations, since if rare variants are a predominant source of the heritability of complex traits, then we will be unlikely to replicate many associations across populations. However, we find that even for those mutations that are incredibly rare (e.g., private to a single chromosome out of 10,000), there can still be a surprisingly broad range of ages, with 9% of these variants being greater than 3,000 years old.

*The age distribution of deleterious versus neutral variants*

Identifying methods for distinguishing variants with strong effects from those with weak effects is critically important. A classic result in population genetics suggests that for variants at the same population frequency, deleterious alleles will on average be younger than neutral alleles [i.e., mutant alleles driven by natural selection will have arisen more recently in the past (*33*)]. This result suggests that if we were able to accurately estimate the ages of mutations across the genome, then we would be able to distinguish deleterious alleles from neutral ones at the same frequency. To evaluate this hypothesis, we rely on simulations where the true age of an allele is known.

In Figure 5 we show the average age of variants in four fitness effect classes across a range of allele frequencies. Note that synonymous variants are neutral, and non-synonymous variants are only exposed to deleterious fitness effects so only the absolute value of the selection coefficient is shown. We find no distinction in the average age of variants that are synonymous, nearly neutral, or even weakly deleterious for any frequency range. In contrast, there is a clear reduction in the average age of the most deleterious variants for common alleles (>1%), and a moderate reduction in age for the low-frequency variants (0.5%-1%) and singleton class (Figure 5).

Since strongly deleterious alleles are on average younger than neutral alleles for low frequency and common variants, we next set out to characterize the distribution of ages for these two categories. Figure 6 shows the average age for neutral (blue; $s > -10^{-5}$) and strongly deleterious (red; $s < -10^{-2}$) mutations as a function of allele frequency. Each curve is contained in an envelope representing the 90% quantile range of all variants observed at that frequency. We find a marked difference in the average age of deleterious and nearly neutral variants that have frequency >1%. The inset in Figure 6 zooms into the <2% allele frequency range, where it is apparent that after 0.5% frequency, the age distribution of neutral alleles starts to widen.  While the age distribution of deleterious variants is largely contained within the distribution of neutral variants, it is clear that a large fraction of neutral variants could potentially be identified based on their age (and therefore eliminated from further follow-up).

**Discussion**

Natural selection is the evolutionary force that acts to either increase or decrease the frequency of a genetic variant based on its effect on reproductive fitness. The effects of natural selection are amplified by demographic changes, with the strength of selection being proportional to population size (*28, 34, 35*). For example, selection has a much weaker effect on allele frequency in smaller populations.

We find that genes associated with several complex disease/trait categories display an excess of negative selection, even in healthy individuals (Tables 1 and S1).  It is plausible that the disease-associated genes showing most evidence for negative selection would also contain a greater excess of rare variants, particularly in patients with the associated disease.  In an application of this hypothesis, nine of the asthma-associated genes showing the strongest signatures of negative selection were sequenced in 450 asthmatic cases and 515 controls from two populations (*36*).  Multiple genes were found to exhibit an excess of rare variants in cases compared to controls.  Broader sequencing efforts are underway to fully evaluate the usefulness of natural selection as a predictor of functionally relevant patterns of genetic variation.

More broadly, we observe that genomic loci associated with complex diseases and traits through GWAS exhibit enriched signals of background selection compared to control SNPs (Figure 1). This implies that GWAS SNPs tend to be located in regions of the genome that are surrounded by loci that are recurrent targets of negative selection. Yet, we find that the SNPs so far identified do not exhibit greater evolutionary constraint than control SNPs. This suggests that targeted sequencing studies around GWAS hits in large samples could be fruitful in discovering causal rare variants.

While GWAS SNPs themselves are not enriched for greater evolutionary constraint, we do observe that genetic diversity across three human populations is highly correlated with long-term evolutionary constraint (Figure 2). However, it remains unclear how much of this correlation is due to ongoing negative selection as opposed to complex patterns of neutral variation. It was previously shown that some of the correlation between conservation and diversity is innate to neutral evolutionary processes, and can be recapitulated in neutral simulations that by chance have lower divergence (*37*). Nonetheless, regions of the genome with the highest level of evolutionary conservation have been found to exhibit greater evidence of negative selection than predicted by neutral evolutionary models (*37*). We do not replicate this result, as we find slightly higher levels of diversity in regions with extremely large phyloP scores (i.e. highly conserved regions) than we predicted based on an extrapolation of the correlation between diversity and divergence that was observed in nearly neutral regions of the genome. At least part of this can be explained by changes in the fitness landscape in recent human evolution, whereby some highly conserved regions have become extremely rapidly evolving (*38*).

Another complication is that the complex demographic history of many human populations (multiple bottlenecks followed by recent explosive growth) has radically shaped patterns of genetic diversity across the genome. During historical bottlenecks, deleterious mutations were tolerated more readily, allowing them to segregate at slightly higher frequencies [and resulting in increased homozygosity of deleterious mutations in some populations compared to others (*32*)]. The explosive growth that humans have experienced over the last ~5,000 years is again reshaping the landscape of natural selection in human populations.

One consequence of the action of natural selection is that most of the strongly deleterious variants are purged from the population before they reach 0.1% frequency (Figure 3). If effect sizes of causal variants for complex disease are correlated with fitness effects, then this is a population genetic justification for studying rare genetic variation across very large population samples. That is, common variation is largely neutral and unlikely to harbor functionally relevant variation. This agrees with many GWAS that have found common variants to generally have very small effect sizes. Having accepted the importance of rare variation, however, the key question then becomes how to distinguish functionally relevant variants (which are likely deleterious) from those that are neutral and ostensibly extraneous.

In Figure 6, we describe the variation in the ages of neutral and strongly deleterious alleles. We find marked differences in the distributions of alleles that have frequency >0.5%. The primary observation is that neutral alleles tend to have a considerably wider distribution of allele ages compared to deleterious alleles, which have a remarkably small variance in age across the entire allele frequency range. This observation suggests a simple criterion for prioritizing variants to be functionally characterized: all else being equal, the youngest alleles in each frequency range should be examined first.

A recent publication described the ages of alleles observed in sample of >6000 exomes (*8*). They point out that the variants that are identified as deleterious by multiple methods tend to be younger on average than predicted neutral variants. However, the method used to date the age of each mutation was based solely on allele frequency, similar to our presentation in Figure 4. One implication is that such a method would not predict a difference in allele age for neutral and deleterious variants at any given frequency (such as we observe in Figure 5). This emphasizes the need for novel methods that utilize more information, such as haplotype lengths and population sharing, for characterizing the ages of alleles.

Nonetheless, population geneticists will still face challenges when developing methods that classify variants as deleterious or neutral, at least for the class of rare variants that are potentially of most interest for large-scale genome sequencing efforts of complex diseases (0.02%-0.5% frequency). In contrast to more common variants (>0.5% frequency), there is no difference in the average age across any fitness class (Figure 5). This counterintuitive result appears to be at odds with population genetic theory (*33*), but is a result of the substantial—and frequently neglected—impact of a complex demographic history. It has previously been shown that population bottlenecks can increase the relative proportion of deleterious variants compared to populations that have not experienced a previous bottleneck (*32*). Moreover, a previous simulation study found that strongly deleterious variants at 1% frequency were no younger than neutral variants in an extremely rapidly growing population (*39*). It is therefore imperative that further studies exhaustively evaluate the population characteristics of rare deleterious variants to provide insight into what distinguishes them from neutral variants.

Among low frequency and common variants (i.e., >0.5% frequency), deleterious alleles are on average younger than neutral alleles. As a result, we expect that these two classes of variants will exhibit very different patterns of haplotype lengths. While the simulations performed in this study are not sufficient to characterize the distribution of haplotype lengths (which requires the simulation of sequences much longer than 1.5kb), we would expect younger alleles to reside on longer haplotypes. The length of a haplotype is directly related to our ability to accurately impute missing data on that haplotype (*40*), which implies that if deleterious alleles are indeed younger on average than neutral variants, then we would expect to have much greater power to accurately impute them.

Increased sequencing efforts will be indispensible for elucidating the role of rare variation in diseases and complex traits.  As the field of human genetics progresses toward the sequencing of more and more human genomes, thorough analyses of genome diversity will continue to be imperative.  We suggest that a careful consideration of the evolutionary processes that shape human genetic variation should play an important role in illuminating the genetic basis of human phenotypic variation.  By combining information from multiple sources and statistical techniques developed across multiple fields, the future is bright for uncovering the genetic architecture of complex traits.

**Acknowledgements**

We thank two reviewers for thoughtful comments on this manuscript.  MCM was supported by the Epidemiology and Translational Science program at UCSF.  LHU was partially supported by NIH training grant T32GM008155 and an ARCS fellowship.  This project was partially supported by NIH/NCRR UCSF-CTSI Grant Number UL1 RR024131 to RDH and by the National Institute On Minority Health And Health Disparities of the National Institutes of Health under Award Number P60MD006902.

**Tables:**

Table 1:  Negative selection operates on genes and conserved non-coding (CNCs) regions near genes that are associated with complex diseases.  Each cell contains the number of complex disease categories (out of 17) with statistically more negative selection compared to background (and binomial-based p-value for enrichment given the number of diseases tested at the 5%

significance level).

|        | AA          | CA          | Pooled      |
|--------|-------------|-------------|-------------|
| CDS    | 3 (0.009)   | 2 (0.05)    | 5 (0.006)   |
| CNCs   | 1 (0.21)    | 3 (0.009)   | 4 (0.03)    |
| Pooled | 4 (0.026)   | 5 (0.006)   | 9 (0.002)   |

**FIGURE LEGENDS**:

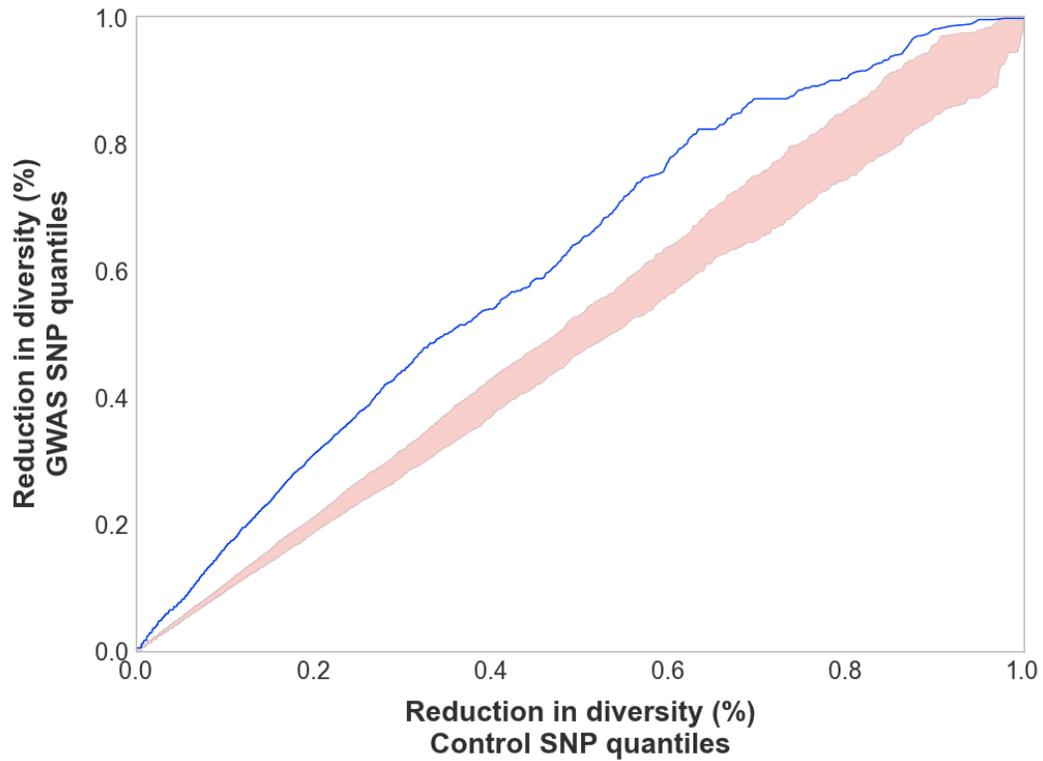

Figure 1. GWAS SNPs for complex traits are significantly more impacted by background selection than are control SNPs in HapMap (*41*) (chosen to match the frequency and genetic distance to closest gene of GWAS SNPs). In blue is a QQ-plot comparing the background selection scores for GWAS quantiles versus the background selection score for control SNPs. In pink is the entire range of background scores for each quantile across 1000 independently drawn control sets.

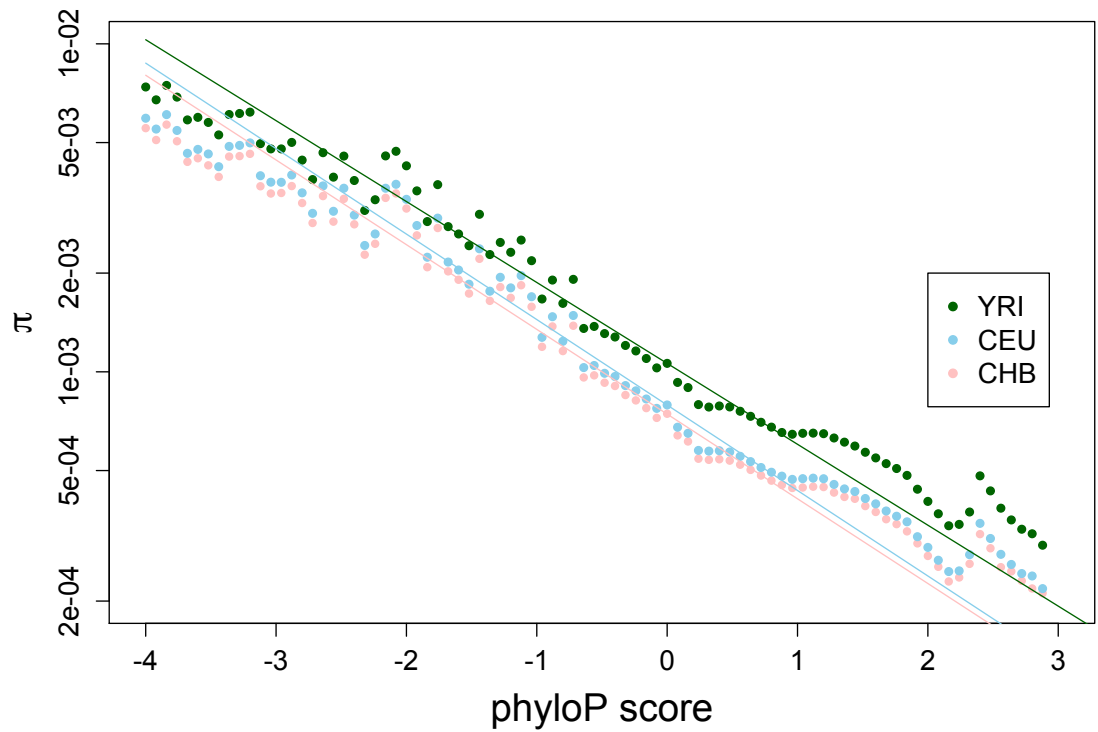

Figure 2: There is a strong log-linear correlation between genetic diversity within three human populations [π; (*7*)] and long-term evolutionary conservation across species [phyloP (*42*)]. Regression lines for each population have been fit to the subset of data considered neutral (phyloP between -1.2 and 1.2), and extrapolated to the range of data observed across the genome.

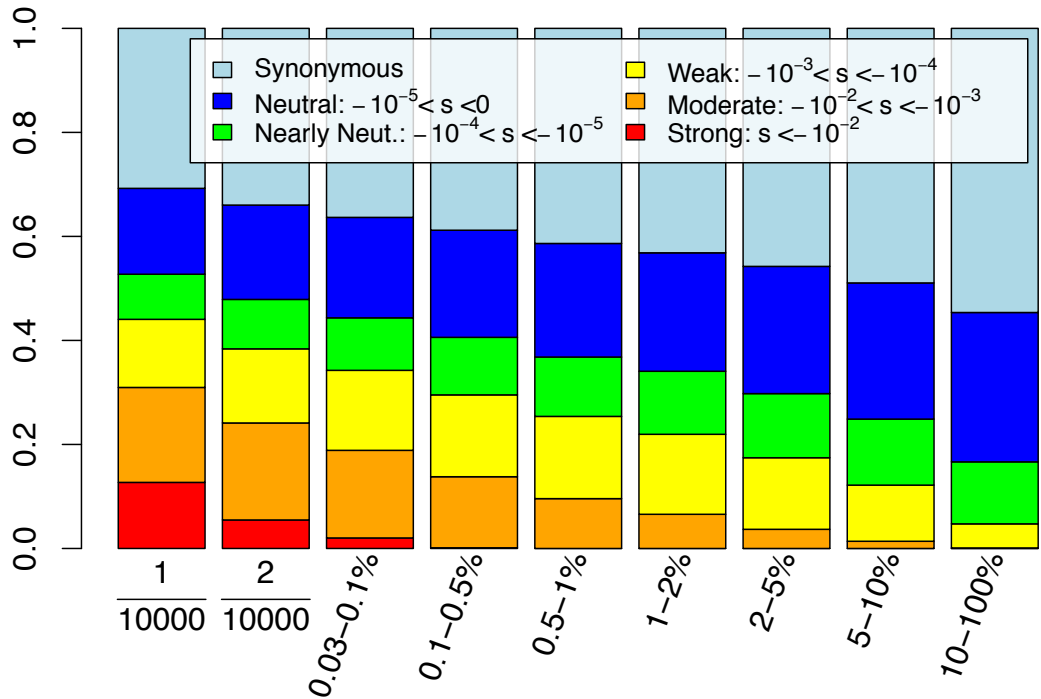

Figure 3. The distribution of selection coefficients for variants in each allele frequency bin (starting with singletons, doubletons, and then a disjoint partition of rare and common variants). Common variants (>5%) are expected to be primarily neutral or nearly-neutral. Most strongly deleterious mutations are <0.1% frequency.

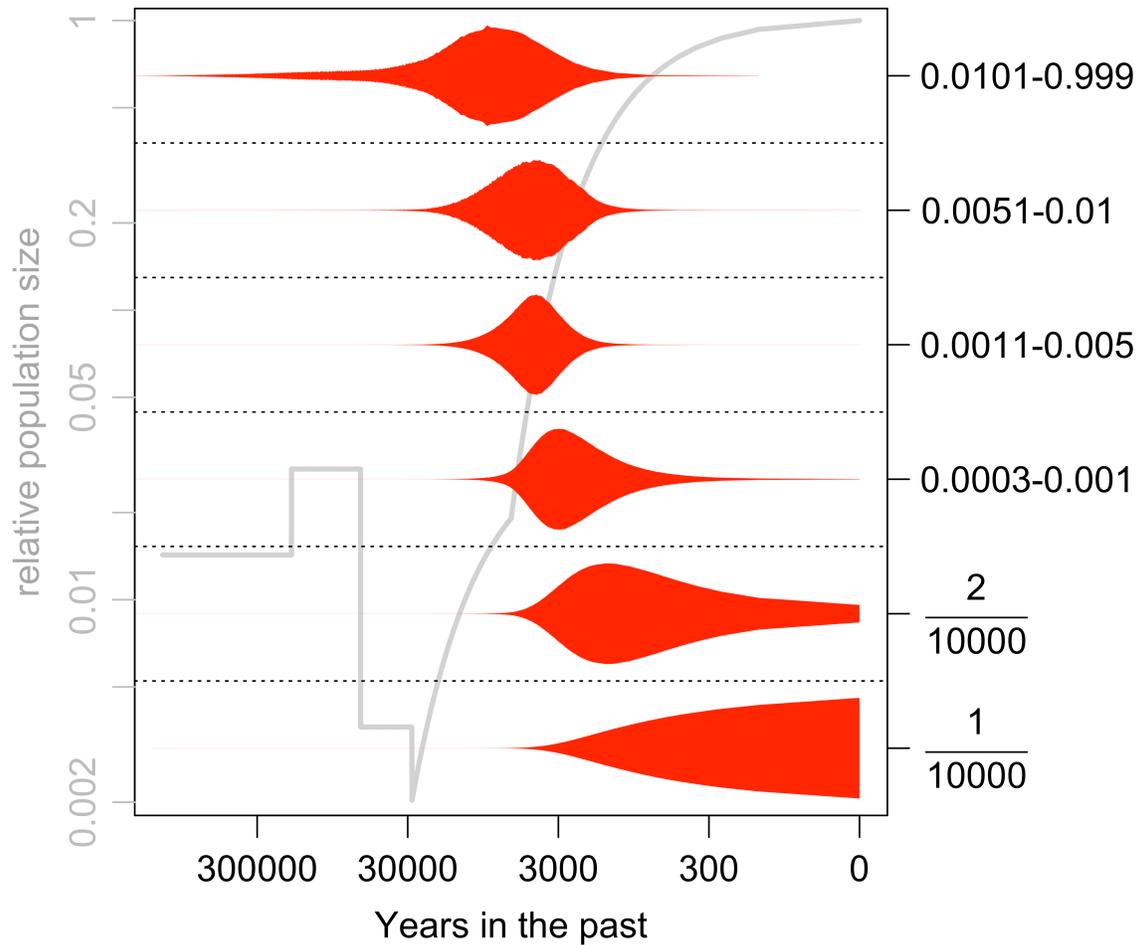

Figure 4. Simulations demonstrate that the patterns of genetic variation of European populations have been shaped by a complex demographic history and distribution of deleterious selective effects. This violin plot shows the distribution of ages of non-synonymous mutations for each allele frequency bin (right axis, including singletons and doubletons, followed by a partitioning of rare variant frequencies), overlaid on top of a model of European demographic history (*9*) (grey, with relative population size on left axis).

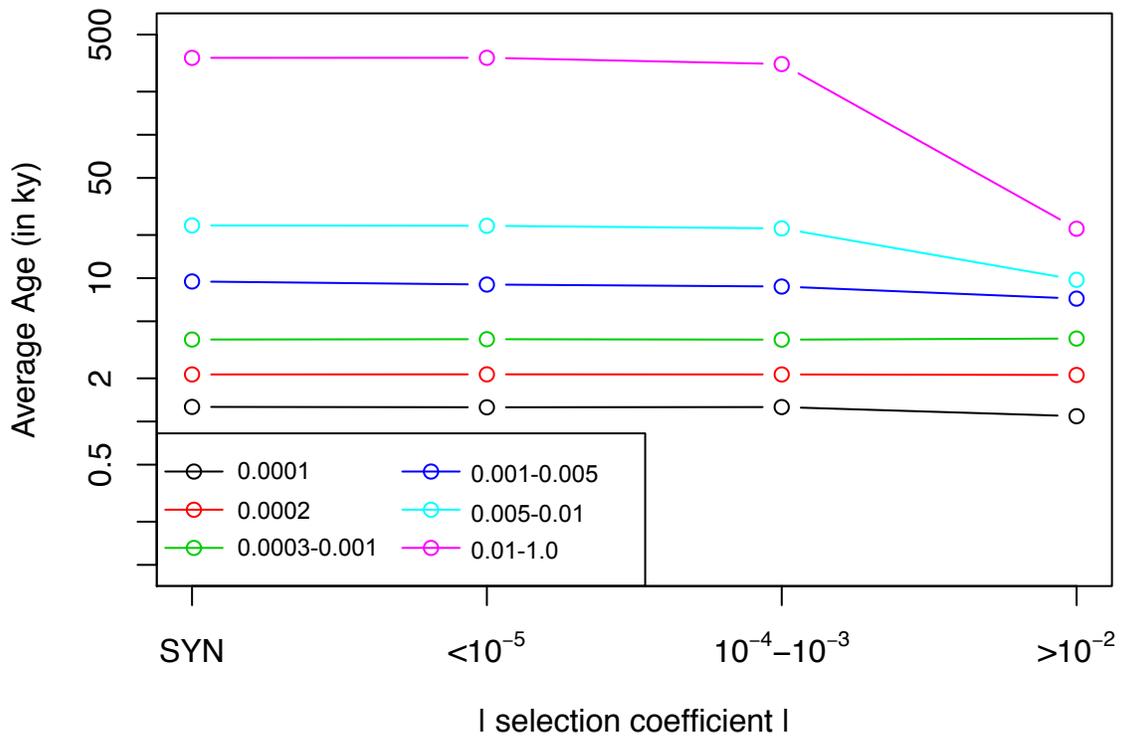

Figure 5. The average age of mutations in various frequency bins, partitioned into synonymous variants, neutral non-synonymous variants ($|s|<10^{-5}$), weakly deleterious variants ($10^{-4}<|s|<10^{-3}$), and strongly deleterious variants ($|s|>10^{-2}$).

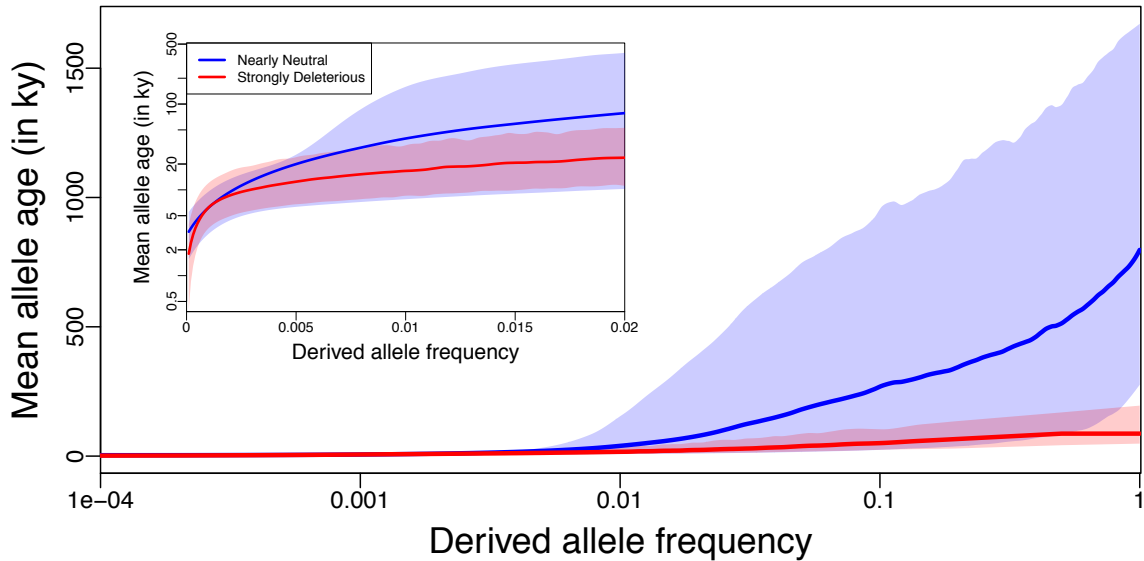

Figure 6. Deleterious alleles are younger on average than neutral alleles at the same frequency. Solid curves represent mean age of alleles at each frequency for non-synonymous variants that are neutral (blue) and strongly deleterious (red). The pink and blue envelopes show the 90%-quantile range of ages for all variants observed at each frequency. Inset is a zoom into the low frequency range showing the increased variance in age for neutral variants at frequency >0.5%. Variants at higher frequencies have been pooled such that the mean and 90%-quantile range are based on at least 200 variants. Curves are smoothed with loess using a span of 0.1.

**SUPPLEMENTAL**

**Table S1:** Negative selection on disease categories. Each disease category is represented by four lines of data. Each line has the population (African American/European American), annotation (conserved non-coding [CNC]/coding sequence [CDS]), the number of genes associated with that disease category for which we have data, the number of genes excluding all disease categories for which we have data, the mean probability of negative selection across genes associated with the disease category, the mean probability of negative selection across genes excluding all disease categories, and the Mann-Whitney U p-value comparing the distributions of the probability of negative selection across genes associated with the disease category versus other genes. P-values that are significant at the 0.05 level are in bold.

|  | pop | data | #Assoc. | Other | meanAssoc | meanOther | p-value |
|---|---|---|---|---|---|---|---|
| Aging | AA | CNC | 24 | 7388 | 0.3265958 | 0.3237955 | 0.2325 |
|  | AA | CDS | 23 | 6358 | 0.5557435 | 0.5191416 | 0.1582 |
|  | CA | CNC | 24 | 7155 | 0.3305875 | 0.329751 | 0.2628 |
|  | CA | CDS | 25 | 6160 | 0.611008 | 0.5676958 | 0.1655 |
| Cancer | **AA** | **CNC** | **173** | **7388** | **0.3394023** | **0.3237955** | **0.04529** |
|  | AA | CDS | 179 | 6358 | 0.5228251 | 0.5191416 | 0.4443 |
|  | **CA** | **CNC** | **167** | **7155** | **0.3636581** | **0.329751** | **0.00352** |
|  | CA | CDS | 180 | 6160 | 0.5773372 | 0.5676958 | 0.298 |
| Cardiovascular | AA | CNC | 201 | 7388 | 0.3282109 | 0.3237955 | 0.358 |
|  | **AA** | **CDS** | **219** | **6358** | **0.549905** | **0.5191416** | **0.02862** |
|  | CA | CNC | 193 | 7155 | 0.3330347 | 0.329751 | 0.5163 |
|  | CA | CDS | 220 | 6160 | 0.5893723 | 0.5676958 | 0.09786 |
| Chemical Dependency | AA | CNC | 41 | 7388 | 0.3458439 | 0.3237955 | 0.109 |
|  | AA | CDS | 30 | 6358 | 0.4423467 | 0.5191416 | 0.9755 |
|  | CA | CNC | 38 | 7155 | 0.3224895 | 0.329751 | 0.4995 |
|  | CA | CDS | 31 | 6160 | 0.5411226 | 0.5676958 | 0.766 |
| Developmental | AA | CNC | 48 | 7388 | 0.3366521 | 0.3237955 | 0.1805 |
|  | AA | CDS | 38 | 6358 | 0.5402158 | 0.5191416 | 0.2671 |

| | | | | | | |
|---|---|---|---|---|---|---|
| | CA | CNC | 47 | 7155 | 0.3592085 | 0.329751 | 0.1082 |
| | **CA** | **CDS** | **42** | **6160** | **0.6416595** | **0.5676958** | **0.00902** |
| Hematological | AA | CNC | 24 | 7388 | 0.3427 | 0.3237955 | 0.09617 |
| | AA | CDS | 32 | 6358 | 0.46665 | 0.5191416 | 0.9219 |
| | CA | CNC | 20 | 7155 | 0.316215 | 0.329751 | 0.8341 |
| | CA | CDS | 31 | 6160 | 0.4790323 | 0.5676958 | 0.9956 |
| Immune | AA | CNC | 169 | 7388 | 0.3211533 | 0.3237955 | 0.3441 |
| | AA | CDS | 198 | 6358 | 0.5317793 | 0.5191416 | 0.1912 |
| | CA | CNC | 165 | 7155 | 0.3414776 | 0.329751 | 0.05543 |
| | CA | CDS | 192 | 6160 | 0.5686474 | 0.5676958 | 0.5519 |
| Infection | AA | CNC | 49 | 7388 | 0.3055 | 0.3237955 | 0.574 |
| | AA | CDS | 64 | 6358 | 0.5347484 | 0.5191416 | 0.303 |
| | CA | CNC | 49 | 7155 | 0.3555531 | 0.329751 | 0.07868 |
| | CA | CDS | 61 | 6160 | 0.5717836 | 0.5676958 | 0.5154 |
| **Metabolic** | AA | CNC | 237 | 7388 | 0.3333046 | 0.3237955 | 0.1513 |
| | **AA** | **CDS** | **249** | **6358** | **0.5379739** | **0.5191416** | **0.04641** |
| | **CA** | **CNC** | **231** | **7155** | **0.3581805** | **0.329751** | **0.0161** |
| | **CA** | **CDS** | **245** | **6160** | **0.5910837** | **0.5676958** | **0.03799** |
| Neurological | AA | CNC | 120 | 7388 | 0.3113 | 0.3237955 | 0.8659 |
| | AA | CDS | 100 | 6358 | 0.518476 | 0.5191416 | 0.5738 |
| | CA | CNC | 113 | 7155 | 0.3291496 | 0.329751 | 0.6071 |
| | CA | CDS | 97 | 6160 | 0.5722299 | 0.5676958 | 0.5123 |
| **Other** | AA | CNC | 219 | 7388 | 0.3244265 | 0.3237955 | 0.4513 |
| | **AA** | **CDS** | **225** | **6358** | **0.5389133** | **0.5191416** | **0.04889** |
| | CA | CNC | 206 | 7155 | 0.3275316 | 0.329751 | 0.3597 |
| | CA | CDS | 219 | 6160 | 0.5750078 | 0.5676958 | 0.2851 |
| **Pharmacogenomic** | AA | CNC | 56 | 7388 | 0.3357036 | 0.3237955 | 0.2611 |
| | AA | CDS | 58 | 6358 | 0.5280845 | 0.5191416 | 0.3902 |
| | **CA** | **CNC** | **50** | **7155** | **0.368852** | **0.329751** | **0.01569** |
| | CA | CDS | 57 | 6160 | 0.5720702 | 0.5676958 | 0.496 |
| Psychological | AA | CNC | 153 | 7388 | 0.3149144 | 0.3237955 | 0.6817 |
| | AA | CDS | 112 | 6358 | 0.526833 | 0.5191416 | 0.3743 |
| | CA | CNC | 151 | 7155 | 0.3359338 | 0.329751 | 0.4872 |
| | CA | CDS | 116 | 6160 | 0.5834086 | 0.5676958 | 0.3154 |
| Renal | AA | CNC | 57 | 7388 | 0.3331719 | 0.3237955 | 0.1028 |
| | AA | CDS | 68 | 6358 | 0.5224221 | 0.5191416 | 0.4384 |
| | CA | CNC | 54 | 7155 | 0.3232741 | 0.329751 | 0.5621 |
| | CA | CDS | 66 | 6160 | 0.5691076 | 0.5676958 | 0.436 |
| Reproduction | AA | CNC | 57 | 7388 | 0.3228404 | 0.3237955 | 0.1809 |
| | AA | CDS | 69 | 6358 | 0.5317464 | 0.5191416 | 0.3314 |
| | CA | CNC | 56 | 7155 | 0.3540446 | 0.329751 | 0.1071 |
| | CA | CDS | 68 | 6160 | 0.5800721 | 0.5676958 | 0.3279 |
| Unknown | AA | CNC | 47 | 7388 | 0.3103128 | 0.3237955 | 0.5766 |
| | AA | CDS | 46 | 6358 | 0.5116196 | 0.5191416 | 0.6982 |
| | CA | CNC | 43 | 7155 | 0.3425163 | 0.329751 | 0.1793 |
| | CA | CDS | 49 | 6160 | 0.5850857 | 0.5676958 | 0.3852 |
| Vision | AA | CNC | 32 | 7388 | 0.3017906 | 0.3237955 | 0.6987 |
| | AA | CDS | 36 | 6358 | 0.4809944 | 0.5191416 | 0.8748 |
| | CA | CNC | 30 | 7155 | 0.33131 | 0.329751 | 0.3997 |
| | CA | CDS | 37 | 6160 | 0.5669081 | 0.5676958 | 0.5373 |

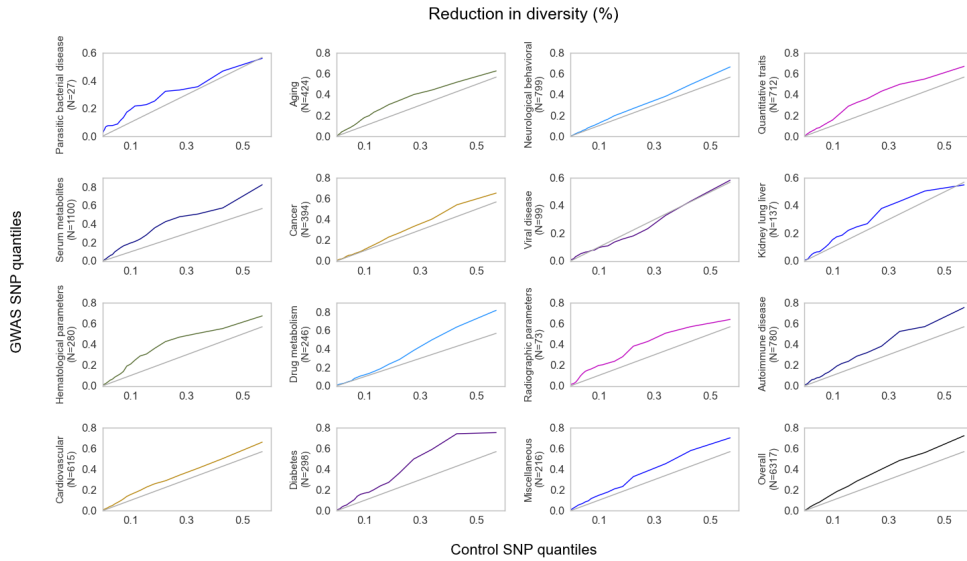

Figure S1. QQ-plots for the effect of background selection in each of 15 complex disease and trait categories (and overall). For each panel, an independent set of 1000 controls sets of SNPs were created to reflect the number of SNPs associated with that particular category as well as the frequency of those alleles and their distance to nearest gene. To provide even smoothness across sample sizes, we have discretized the plot into five percentile intervals (vigintiles).